\documentclass[prd,showpacs,letterpaper,twocolumn]{revtex4}%
\usepackage{graphicx}
\usepackage{bm}
\usepackage{epsf}
\usepackage{rotating}
\usepackage{epsfig,graphics,rotate,color}
\usepackage{wrapfig}
\usepackage{amssymb}
\usepackage{amsmath}
\usepackage{amsfonts}
\usepackage{array,hhline,dcolumn}%
\providecommand{\U}[1]{\protect\rule{.1in}{.1in}}
\begin{document}
\title{A New Method to Search for $CP$ violation in the Neutrino Sector}
\author{J.M. Conrad$^1$ and M.H. Shaevitz$^2$}
\affiliation{$^{1}$ Massachusetts Institute of Technology}
\affiliation{$^{2}$ Columbia University}

\begin{abstract}
  New low-cost, high-power proton cyclotrons open the opportunity for
  a novel precision search for $CP$ violation in the neutrino
  sector. The accelerators can produce decay-at-rest neutrino beams
  located at multiple distances from a Gd-doped ultra-large water
  Cerenkov detector in order to search for $CP$ violation in $\bar
  \nu_\mu \rightarrow \bar \nu_e$ at short baseline. This new type of
  search complements presently proposed experiments, providing
  measurements that could lead to a substantially better exploration of $CP$
  violation in the neutrino sector.

\end{abstract}
\maketitle

\section{Introduction}

With the discovery of neutrino oscillations, particle physicists have
been inspired to develop theories that explain very light neutrino
masses.  The most popular models invokes GUT-scale Majorana
partners which can decay, producing a matter-antimatter asymmetry in
the early universe through the mechanism of $CP$ violation.  Observation
of $CP$ violation in the light neutrino sector would be a strong hint
that this theory is correct.

To incorporate $CP$ violation, the light-neutrino 
mixing matrix is expanded to include a $CP$ violating phase,
$\delta_{CP}$. Sensitivity to  $\delta_{CP}$ comes 
through muon-to-electron flavor oscillations at the atmospheric
mass-squared difference, $\Delta m_{31}^{2}$. The oscillation probability, 
neglecting matter effects, is given by \cite{Parke}:
\begin{align}
P  & =\sin^{2}\theta_{23}\sin^{2}2\theta_{13}\sin^{2}%
\Delta_{13} \nonumber \\
& \mp\sin\delta_{cp}\sin2\theta_{13} \sin2\theta_{23}\sin2\theta _{12} \nonumber \\
& ~~~~~~~~~~~~~~~\times \sin^{2}\Delta_{13}\sin\Delta_{12} \nonumber \\
& +\cos\delta_{cp}\sin2\theta_{13} \sin2\theta_{23} \sin2\theta 
_{12} \nonumber \\
&  ~~~~~~~~~~~~~~~\times\sin\Delta_{13}\cos\Delta_{13}\sin\Delta_{12} \nonumber \\
& +\cos^{2}\theta_{23}\sin^{2}2\theta_{12}\sin^{2}\Delta_{12},\label{equ:beam}
\end{align}
where $\Delta_{ij}=\Delta m_{ij}^{2}L/4E_{\nu}$, and 
$-(+)$ refers to neutrinos (antineutrinos).

In Eq.~\ref{equ:beam}, aside from $\delta_{CP}$, all but two of the
parameters ($\theta_{13}$ and the sign of $\Delta m_{31}^2$) are well
known.  Further precision on these known parameters is expected in the
near future (see Table~\ref{OscPar}).  With respect to $\theta_{13}$,
global fits report a non-zero value at the $\sim 1\sigma$ level
\cite{schwetz, fogli}.  This parameter drives the amplitude for the $CP$
violating terms in Eq.~\ref{equ:beam} and therefore sets the level of
technical difficulty for observing $CP$ violation.  The unknown sign of
$\Delta m_{31}^2$, referred to as ``the mass hierarchy,'' affects the
sign of term 3 in Eq.~\ref{equ:beam}.  Matter effects modify 
Eq.~\ref{equ:beam} and are sensitive to the mass hierarchy.
  
We propose a new method to search for $CP$ violation by comparing
absolute neutrino rates in a single detector that is illuminated by
neutrino beam sources at multiple distances.  This method explores
the $\delta_{CP}$ dependence inherent in Eq.~\ref{equ:beam} in
a different way and with comparable precision than the planned
long-baseline program.

The neutrino sources would be based on
commercially-developed, small (2.5 m diameter), high-power proton
cyclotron accelerators that are under development~\cite{Timothy}. A 250
MeV, 1 mA cyclotron is under construction at MIT and a GeV-energy,
megawatt-class cyclotron is presently under design.  When in production,
because of new, inexpensive superconducting technology, these machines
are expected to cost 5\% of a conventional proton accelerator ($<
\$20{\rm M}$).  These machines will demonstrate the beam physics and
engineering needed for this experiment.

\begin{figure}[b]\begin{center}
{\includegraphics[width=3.5in]{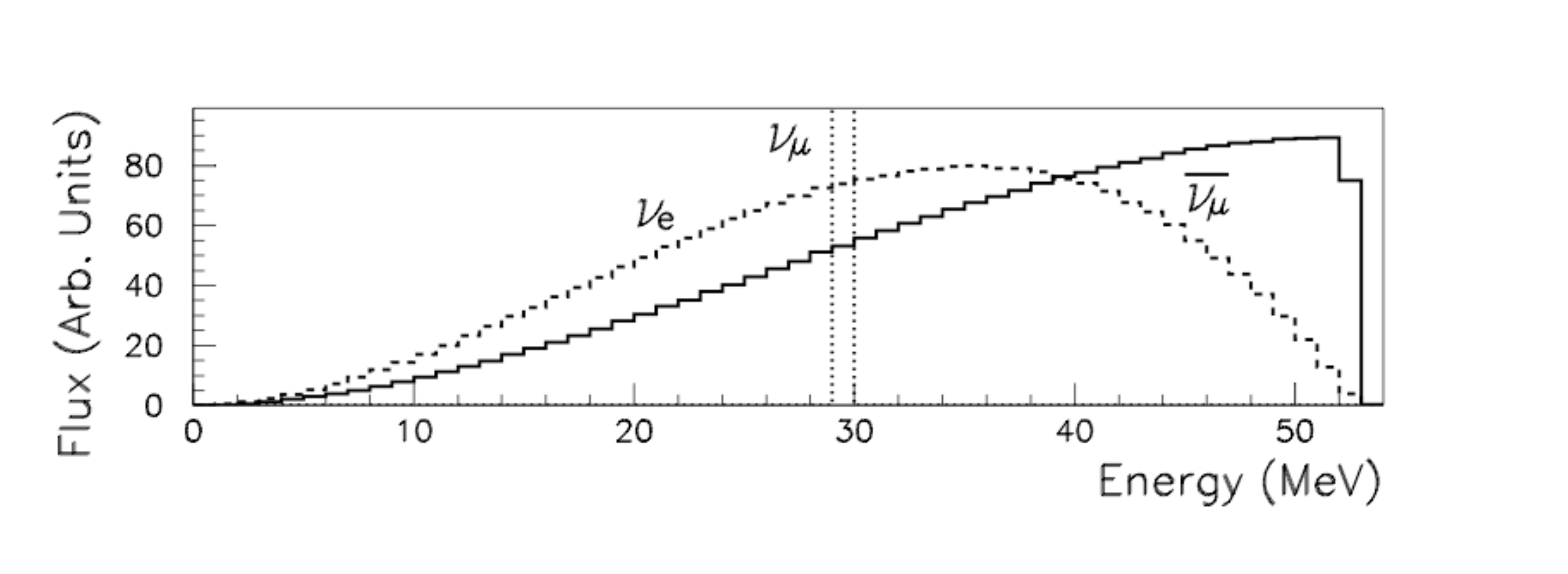}
} \end{center}
\vspace{-0.25in}
\caption{Energy distribution of neutrinos in a DAR beam 
\label{fluxplot} }
\end{figure}

This experiment requires accelerators that target 2 GeV protons at 2.5 mA
during a 100 $\mu$s pulse every 500 $\mu$s, delivering
$9.4\times10^{22}$ protons per year to a beam stop. The result is a
high-intensity, isotropic, decay-at-rest (DAR) neutrino beam 
arising from the stopped pion decay chain:
$\pi^{+}\rightarrow\nu_{\mu}+ \mu^{+}$ followed by $\mu^+ \rightarrow
e^{+}\bar{\nu}_{\mu}\nu_{e}$.  The flux, shown in 
Fig.~\ref{fluxplot}, has an endpoint of 52.8 MeV.  Each cyclotron produces
$4\times 10^{22}$/flavor/year of $\nu_e$, $\nu_\mu$ and $\bar \nu_\mu$.
The
$\bar \nu_e$ fraction in the beam is very low ($\sim 10^{-4}$)
because most $\pi^-$ are captured before decay.

This experiment utilizes an ultra-large water Cerenkov detector such
as Hyper-K \cite{HK}, MEMPHYS \cite{MEM}, or the detector proposed for
the Deep Underground Science and Engineering Laboratory (DUSEL)
\cite{LBNE}.  We use a 300 kton DUSEL detector as our model. Water
provides a target of free protons for the inverse beta decay (IBD)
interaction: $\bar \nu_e + p \rightarrow n + e^+$.

IBD interactions are identified via a coincidence signal.  The first
signal is from the Cerenkov ring produced by the positron.  The second
signal is from capture of the neutron. The signal from neutron capture
on protons, which produces only a single 2.2 MeV $\gamma$, is too
feeble to be efficiently observed in a large Cerenkov detector.  
For that reason, doping with gadolinium (Gd), which has a high capture rate and
a short capture time, is
proposed \cite{GADZ}.  Neutron capture on Gd produces multiple photons
totaling $\sim$8 MeV which can be observed with 67\% efficiency
\cite{ntag}.  Gd-doping of water is under development \cite{LLNL}.  In
IBD events, the energy of the neutrino is related to the energy of the
positron and kinetic energy of the neutron, $K$, by
$E_{\overline{\nu}}=E_{e^{+}}+(M_{n}-M_{p})+K$ \cite{BeacomVogel}.

\begin{figure}[t]\begin{center}
{\includegraphics[width=3.25in]{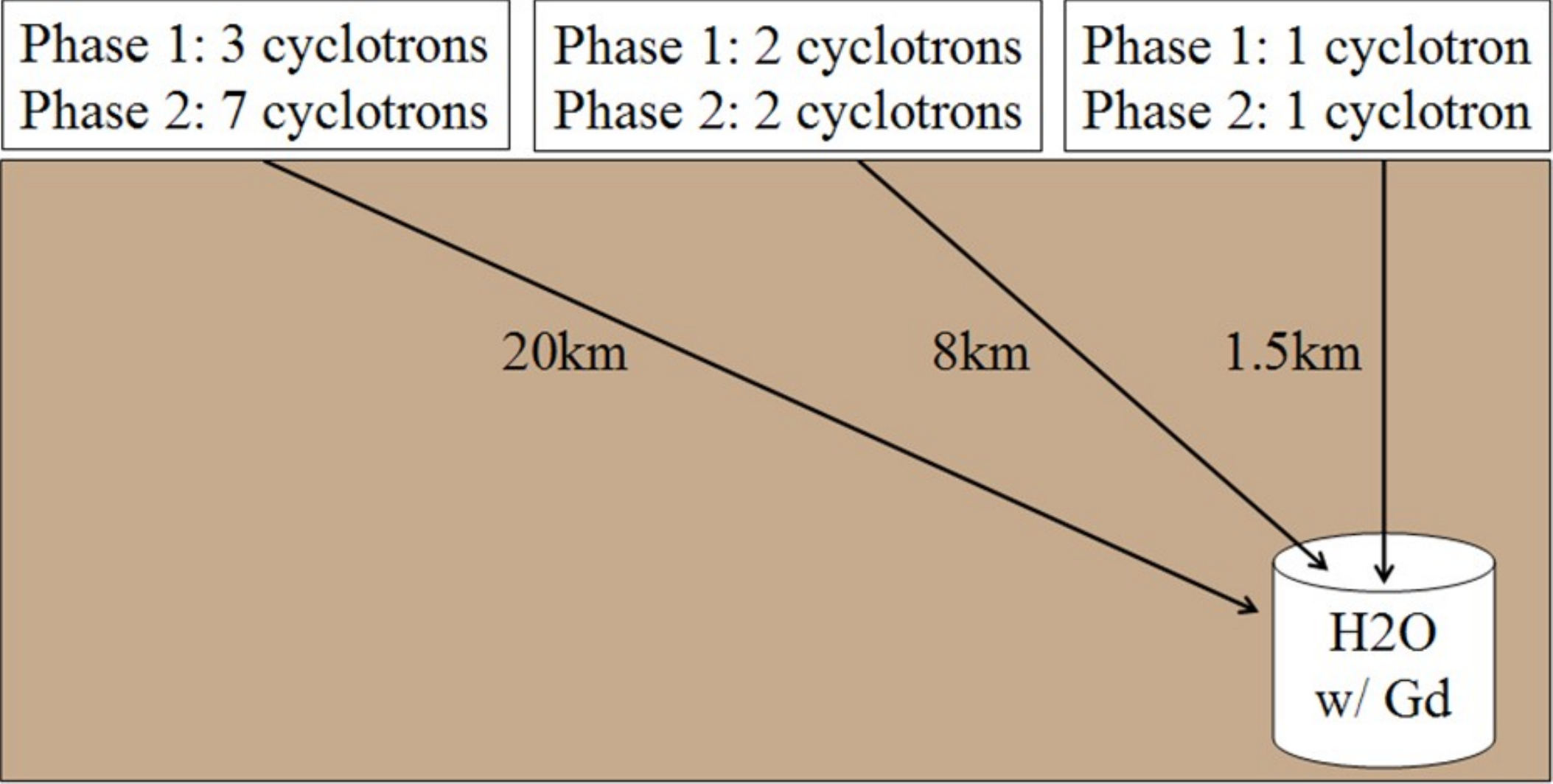}
} \end{center}
\vspace{-0.25in}
\caption{Proposed experimental layout.  Each phase corresponds to a 5-year 
data-taking period with each of the three source sites on for
  20\% of the time.  (Phase 2 is an example; the number of cyclotrons
  will be optimized based on Phase 1 measurements.)
\label{layout} }
\end{figure}

The accelerators are placed at three separate baselines, $L$, for the
$\bar \nu_\mu \rightarrow \bar \nu_e$ search (see Fig.~\ref{layout}). 
Each set of accelerators at the three sites runs for a 20\% 
duty factor interspersed in time; the other 40\% of the time
is used to collect beam-off data. The longest baseline is
$L=20$ km, where $\sim$40 MeV $\bar \nu_\mu$ are at oscillation
maximum for the measured value of $\Delta m^2_{31}$.  In this case, term 2 of
Eq.~\ref{equ:beam} will contribute while term 3 will not.  A
mid-baseline accelerator site is located at $L=8$ km, where term 3 of
Eq.~\ref{equ:beam} is non-zero.  The shortest baseline accelerator site
is at $L=$1.5 km,
directly above the DUSEL detector.  At this location,
the neutrino-electron elastic scattering rates are sufficient to allow
precision measurement of the neutrino flux normalization as discussed
below.  We propose
multiple accelerators at the various locations (see Fig. \ref{layout}), 
which is feasible because of the low cost per machine.
For the $L$ of a given event to be well-determined, the beams from each 
location are staggered in time. 

Term 3 of Eq. 1 leads to an inherent ambiguity between 
$\delta_{CP}$ and the mass hierarchy. We assume that 
on the timescale of the measurements presented here, 
the mass hierarchy will be measured using LBNE\cite{LBNE} 
or atmospheric neutrino measurements.\cite{INO}.  
For the sensitivity estimates, we choose the normal hierarchy 
as the example model.

We present a two-phase experiment, where the design has flexibility in
the second phase.  In our case, each phase represents five years.  The
first phase explores the oscillation space at $\sim$3$\sigma$ and
requires one accelerator at the near location, two at the mid-location
and three at the far location.  Once a signal for $CP$ violation has
been localized, the strategy for the second phase can be determined.
As an example for phase 2, we have chosen a design with one, two and seven
accelerators.  This design matches the sensitivity of a run of LBNE with $30
\times 10^{20}$ protons on target in neutrino mode followed by $30
\times 10^{20}$ protons on target in antineutrino mode \cite{LBNE}.

This design is unique among proposals for $CP$ violation
searches and complementary to the present plans \cite{MikeKendall, LBNE}.
The measurement is done with antineutrinos, while
all existing proposals rely heavily on neutrino data.
The neutrino production systematics are different and well-controlled.
Because of the low energy, the interaction systematics are also
different. Varying $L$, while employing a single detector is
novel and reduces systematics.  
A two-phase program which allows an optimized measurement
strategy is powerful and potentially cost-saving.

%

\begin{table}[tbp] \centering
%
\begin{tabular}
[c]{c|ccc|ccc}\hline
Parameter & Present: &  &  & Assumed & Future:  & \\
& Value& Uncert. & Ref. & Value& Uncert. & Ref.\\
&      & $(\pm)$ &  &          &  $(\pm)$ &   \\ \hline
$\Delta m_{21}^{2}\times 10^{-5} {\rm eV}^{2} $ & 7.65 & 0.23 &
\cite{schwetz} & 7.65 & --- & ---\\

$\Delta m_{31}^{2} \times 10^{-3} {\rm eV}^{2}$ & 2.40 & 0.12 &
\cite{schwetz} & 2.40 & 0.02 & \cite{schwetz05}\\
$\sin^{2}(2\theta_{12})$ & 0.846 & 0.033 & \cite{schwetz} & 0.846 & --- &
---\\
$\sin^{2}(2\theta_{23})$ & 1.00 & 0.02 & \cite{schwetz} & 1.00 & 
0.005 & \cite{huber}\\
$\sin^{2}(2\theta_{13})$ & 0.06 & 0.04 & \cite{fogli} & 0.05 & 
0.005 & \cite{MikeKendall}\\\hline
\end{tabular}
\caption{Left: Present values and uncertainties for oscillation parameters, reported in the
references. Right:  Future expectations used in this study.}\label{OscPar}
\end{table}%

\section{Event Types and Backgrounds}

The energy range of the analysis is $20<E_{\nu}<55$ MeV. The lower cut
renders potential backgrounds from radioactive decay and spallation
negligible.  The upper cut is chosen because of the 52.8 MeV signal endpoint.

In this energy range, three types of interactions must be considered.
First is the IBD signal, with an estimated reconstruction efficiency of
$\epsilon_{recon} = 67\%$, based on studies for Super-K \cite{ntag}.  Second is $\nu_{e}
+ {\rm O} \rightarrow e^{-}+ {\rm F}$ (``$\nu_e$O''), with a suppressed cross section
relative to IBD due to nuclear effects
\cite{Volpe, HaxtonTotal} and no neutron capture.  Third is
neutrino-electron elastic scattering ($\nu_{e}$ES) $\nu_{e} + e^{-}
\rightarrow \nu_{e} + e^{-}$.  This is separable from $\nu_e$O by
angular cuts \cite{LSNDnue, HaxtonAngle} and, so, can be used for the flux
normalization.  For $\nu_e$O and
$\nu_{e}$ES events, we estimate $\epsilon_{recon}=75\%$ \cite{LSNDnue}.

Beam-off backgrounds arise from three sources: atmospheric
$\nu_{\mu}p$ scatters with muons below Cerenkov threshold which stop
and decay (``invisible muons''), atmospheric IBD events, and diffuse
supernova neutrinos. These are all examples of correlated backgrounds;
accidental beam-off backgrounds are estimated to be negligible.  The
rates of these correlated backgrounds are scaled from analyses for the
GADZOOKS experiment \cite{GADZ}.  The interaction rates of the
beam-off backgrounds are well-measured during the 40\% beam-off
running fraction.

The beam-on backgrounds have both accidental and correlated
sources. The accidental backgrounds arise from the $\nu_{e}$ in the
beam which are followed by a neutron-like event.  This
background is estimated to be very small using the measurements 
from the Super-K Gd-doping study\cite{ntag}.  
Correlated backgrounds are produced by the intrinsic
$\bar \nu_{e}$ content of the beam.  This is reduced by careful design
of the beam stop.  We assume a long water target embedded in a copper
absorber, surrounded by steel for each cyclotron.  We target at 100
degrees from the detector to reduce decay-in-flight (DIF)
backgrounds. For 2 GeV protons on target, the pion yields, $\pi^{+}/p$
($\pi^{-}/p$), are 0.43 (0.056). with a $\pi^{+}$ ($\pi^{-}$) DIF
fraction of 0.017 (0.025).  The decays of $\mu^-$'s are further
reduced by a factor of 8.3 since the $\mu^-$ will be captured before
decaying.  The compact design decreases the $\bar{\nu}_{e}$ production
compared to LAMPF/LANCE by a factor of 1.7. The beam has a
$\bar \nu_e/\nu_e$ ratio of $4\times10^{-4}$.

\bigskip%
\begin{table}[tbp] \centering
\begin{tabular}
[c]{l|ccc}\hline
Event Type & 1.5 km & 8 km & 20 km\\\hline
IBD Oscillation Events (E$_{\nu}>20$ MeV) &  &  & \\
$\delta_{CP}=0^{0}$, Normal Hierarchy & 763 & 1270 & 1215\\
\quad\quad" \quad, Inverted Hierarchy & 452 & 820 & 1179\\
$\delta_{CP}=90^{0}$, Normal Hierarchy & 628 & 1220 & 1625\\
\quad\quad" \quad, Inverted Hierarchy & 628 & 1220 & 1642\\
$\delta_{CP}=180^{0}$, Normal Hierarchy & 452 & 818 & 1169\\
\quad\quad" \quad, Inverted Hierarchy & 764 & 1272 & 1225\\
$\delta_{CP}=270^{0}$, Normal Hierarchy & 588 & 870 & 756\\
\quad\quad" \quad, Inverted Hierarchy & 588 & 870 & 766\\\hline
 IBD from Intrinsic $\overline{\nu}_{e}$  (E$_{\nu}>20$ MeV) & 600 & 42 & 17\\
IBD Non-Beam  (E$_{\nu}>20$ MeV) &  &  & \\
\multicolumn{1}{r|}{atmospheric $\nu_{\mu}p$ \textquotedblleft invisible
muons\textquotedblright} & 270 & 270 & 270\\
\multicolumn{1}{r|}{atmospheric IBD} & 55 & 55 & 55\\
\multicolumn{1}{r|}{diffuse SN neutrinos} & 23 & 23 & 23\\\hline
$\nu_{e}-$e Elastic  (E$_{\nu}>10$ MeV) & 16750 & 1178 & 470\\\hline
$\nu_{e}-$Oxygen (E$_{\nu}>20$ MeV) & 101218 & 7116 & 2840\\\hline
\end{tabular}
\caption{Event samples for the combined two-phase run for $\sin^2 2\theta_{13}=0.05$ and
parameters from Table \ref{OscPar} (future). 
}\label{events_123_127}%
\end{table}%

\section{Systematic Errors \label{sys}}

The multiple-DAR-source design is an elegant choice for precision
oscillation physics because the beam and detector systematics are
low. The shape of the DAR flux with energy is known to high precision
and is common among the various distances, thus shape comparisons will
have small uncertainties. The neutrino flux from the three distances
is accurately determined from the direct measurement of the $\pi^{+}$
production rate using $\nu_{e}$ES events from the near accelerator.
Existing proton rate monitors assure that relative proton intensities 
are understood to 0.1\%.  The interaction and detector
systematic errors are low since all events are detected in a single
detector. The IBD cross section for the signal is well-known
\cite{BeacomVogel}.  The fiducial volume error on
the IBD events is also small due to the extreme volume-to-surface-area ratio
of the ultra-large detector.

\begin{figure}[t]\begin{center}
{\includegraphics[width=3.in]{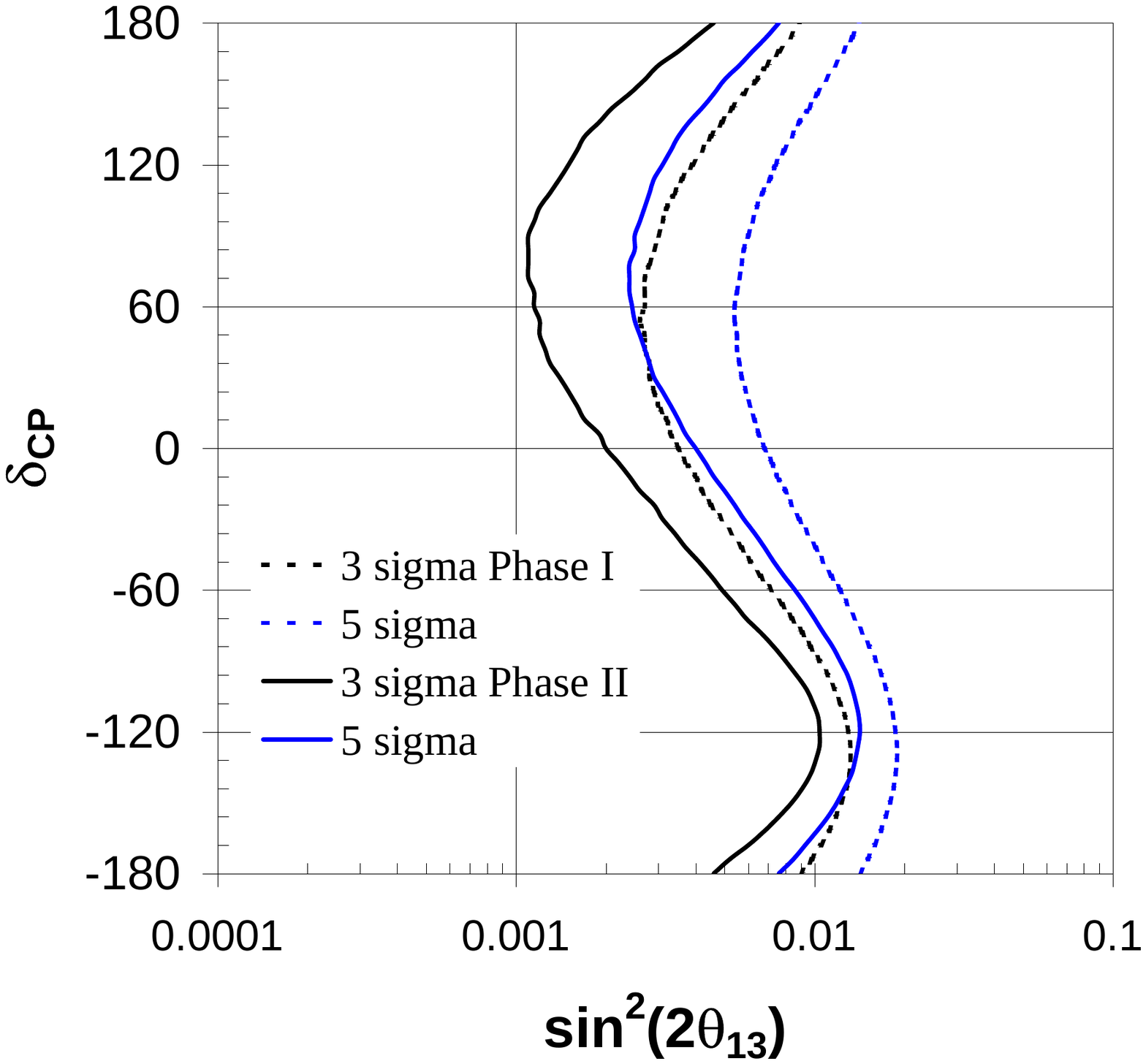}
\vspace{-1.25in}
} \end{center}
\caption{Phase 1 and Combined-phase sensitivity to $\theta_{13} \neq 0$ at 3$\sigma$ and 5$\sigma$.  
\label{thetaplot} }
\vspace{-.75in} \begin{center}
{\includegraphics[width=3.in]{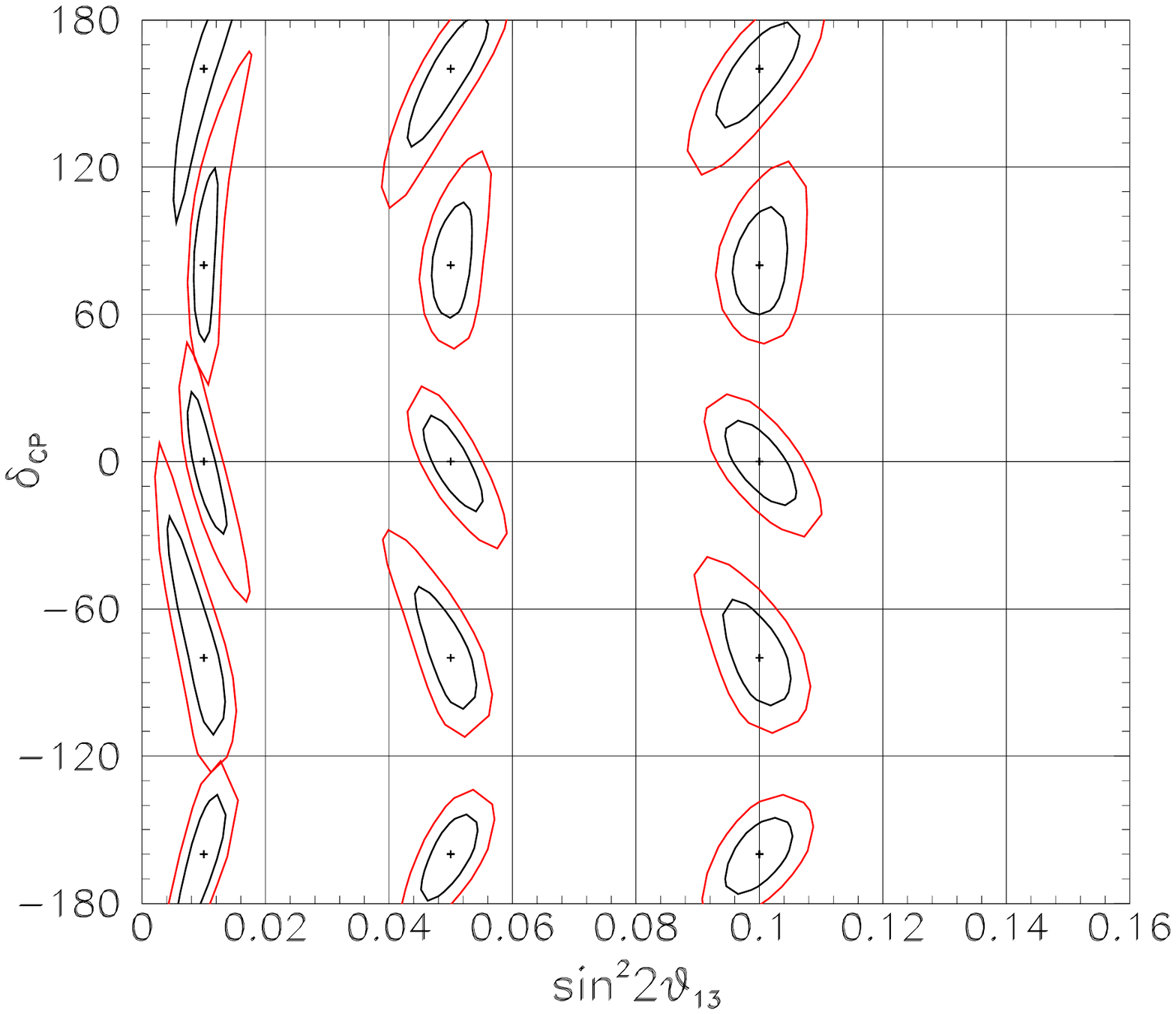}
\vspace{-1.in}
} \end{center}
\caption{Correlated sensitivity (1 and 2$\sigma$ contours) to $\delta_{CP}$ and $\sin^2 2\theta_{13}$ for
the combined two-phase running.
\label{deltaplot} }
\end{figure}

The largest systematic errors arise from the $\nu_{e}$ES 
sample used to determine the absolute normalization of the flux. The
error on this cross section is 0.5\% due to a 0.7\% uncertainty from
the NuTeV $\sin^2 \theta_W$ measurement \cite{sin2thwpaper}.  We assume
a 2.1\% energy scale error \cite{K2KLOI} which leads to a 1\% error on
the DAR flux when a $E_{vis}>10$ MeV cut on the events is applied.  The
$\nu_{e}$ events on oxygen and
IBD events with a missing neutron can be separated from the
$\nu_{e}$ES sample since the angular
distribution of $\nu_{e}$ES events is very forward-peaked, while these
backgrounds have a broad distribution \cite{HaxtonAngle}. 
We take the uncertainty from contamination by these events to be
negligible.  The uncertainty of the electron-to-free-proton
ratio in water is also very small.  Adding the $\nu_{e}$ES systematics in
quadrature, the systematic error on the IBD flux is expected to be
1.1\%. For the total error on the IBD flux, one must then add the
$\nu_{e}$ES statistical error in quadrature, which depends on 
the running period.

The other significant systematic error is on the efficiency for
neutron detection in IBD events. To reduce uncertainties, neutrons are
tagged via timing rather than position reconstruction. This leads to
an inefficiency for neutrons outside of the timing window with a systematic
uncertainty of 0.5\% \cite{ntag}.

\section{Oscillation Sensitivities \label{sec:physics}}

Sensitivity estimates were made by calculating the $\chi^{2}$ for a
given set of predicted events and investigating the $\chi^{2}$
minimization and excursion, using a method similar to
Ref.~\cite{MikeKendall}.  Matter effects were included in the fit, but
are negligible due to the short baseline.  Systematic uncertainties (see
Sec.~\ref{sys} and Table \ref{OscPar}) were constrained by
pull-term contributions to the $\chi^{2}$ of the form $\left(
  k_{i}-1\right)^{2} /\sigma_{i}^{2}$, where $\sigma_{i}^{2}$ are the
uncertainties.  Results are given for two scenarios (see
Fig.~\ref{layout}): Phase 1 and Phase 1+2 combined.  Table
\ref{events_123_127} gives the event samples associated with the
combined phases, for the various classes of oscillation, background,
and calibration events.

The sensitivity for observing a non-zero value for $\theta_{13}$ at the 3 and
5$\sigma$ CL as a function of $\delta_{CP}$ is shown in Fig.
\ref{thetaplot} for combined two-phase running.  This sensitivity meets that 
of LBNE, but is inverted in its $\delta_{CP}$ dependence \cite{LBNE}.

The combined two-phase running yields a 4.1$\sigma$ measurement of
$\delta_{CP}$ at the test point of $\sin^{2}2\theta
_{13}=0.05$ and $\delta_{CP}= -90^{0}$,  as shown in Table
\ref{deltaCP_123_127} (top).  The correlated measurement uncertainties are
shown in Fig.  \ref{deltaplot} for 1 and 2$\sigma$ contours. Table
\ref{deltaCP_123_127} (bottom), which provides the statistical uncertainty,
indicates that the measurement is statistics-limited.

\bigskip%
\begin{table}[tbp] \centering
\begin{tabular}
[c]{c|ccccc}\hline
sin$^{2}2\theta_{13}$/$\delta_{CP}$ & -180 & -90 & 0 & 90 & 135\\\hline
0.01 & 52.5 & 47.2 & 29.0 & 38.8 & 48.5\\
0.05 & 21.6 & 21.9 & 19.5 & 25.6 & 30.4\\
0.09 & 18.3 & 19.9 & 17.6 & 23.8 & 26.3\\\hline 
0.01 & 51.3 & 45.5 & 26.9 & 36.7 & 46.9\\
0.05 & 19.9 & 21.2 & 18.2 & 24.2 & 29.6\\
0.09 & 16.8 & 19.2 & 16.5 & 22.6 & 25.3\\\hline
\end{tabular}
\caption{The 1$\sigma$ measurement uncertainty on $\delta_{CP}$ for various values of
$\sin^{2}\theta_{13}$ for the combined two-phase data. 
Top:  Systematic and statistical errors.  
Bottom: Statistics only.}\label{deltaCP_123_127}%
\end{table}%
%

\section{Conclusions}

We have described a novel experiment to search for $CP$ violation in the
neutrino sector.  This experiment is relatively low cost, employing
multiple high-powered compact cyclotrons located at $L=$ 1.5 km, 8 km and 20 km
from a large water Cerenkov detector. Using the example of the DUSEL
detector, for $\sin^{2}2\theta_{13} = 0.05$, the $CP$ violation
parameter $\delta_{CP}$ can be measured to the level of LBNE ($> 4\sigma$) in
a 2-phase 10-year run.

The complementary nature of this measurement makes it a compelling
addition to the program. This experiment will probe for $CP$ violation
with antineutrinos, unlike the present program which relies heavily on
neutrino interactions. The systematics are also quite different and
low compared to present planned experiments.   As a result, this
experiment will provide a powerful input to our global search for
new physics in the neutrino sector.

\begin{center}
{ {\bf Acknowledgments}}
\end{center}

The authors thank the DAE$\delta$ALUS group for discussions, especially
Timothy Antaya, William Barletta and William Louis.  We thank the
National Science Foundation for support.


\begin{thebibliography}{99}


\bibitem{Parke}
H.~Nunokawa, S.~J.~Parke and J.~W.~F.~Valle,
  Prog.\ Part.\ Nucl.\ Phys.\  {\bf 60}, 338 (2008).
 

\bibitem{schwetz}   T.~Schwetz, M.~A.~Tortola and J.~W.~F.~Valle,
  New J.\ Phys.\  {\bf 10}, 113011 (2008).

\bibitem{fogli}   G.~L.~Fogli, E.~Lisi, A.~Marrone, A.~Palazzo and A.~M.~Rotunno,
  arXiv:0905.3549 [hep-ph].


\bibitem{schwetz05}  T.~Schwetz,
  Acta Phys.\ Polon.\  B {\bf 36}, 3203 (2005).


\bibitem{huber} 
  P.~Huber, M.~Lindner, T.~Schwetz and W.~Winter,
  JHEP {\bf 0911}, 044 (2009).




\bibitem{MikeKendall}  K.~B.~M.~Mahn and M.~H.~Shaevitz,
  Int.\ J.\ Mod.\ Phys.\  A {\bf 21}, 3825 (2006).


\bibitem{Timothy} Private Communication. For further information on the Gigatron, contact T.~Antaya, 
NW22-139, 77 Massachusetts Avenue, Cambridge, Ma 02139.

\bibitem{HK} 
  M.~Aoki, K.~Hagiwara and N.~Okamura,
  Phys.\ Lett.\  B {\bf 554}, 121 (2003).

\bibitem{MEM} 
  A.~de Bellefon {\it et al.},
  arXiv:hep-ex/0607026.

 

\bibitem{LBNE}   V.~Barger {\it et al.},
  arXiv:0705.4396 [hep-ph].



\bibitem{GADZ}   J.~F.~Beacom and M.~R.~Vagins,
  Phys.\ Rev.\ Lett.\  {\bf 93}, 171101 (2004).



\bibitem{ntag}   H.~Watanabe {\it et al.}  [Super-Kamiokande Collaboration],
 arXiv:0811.0735 [hep-ex].


\bibitem{LLNL}   S.~Dazeley, A.~Bernstein, N.~S.~Bowden and R.~Svoboda,
  Nucl.\ Instrum.\ Meth.\  A {\bf 607}, 616 (2009).


\bibitem{BeacomVogel} 
  P.~Vogel and J.~F.~Beacom,
  Phys.\ Rev.\  D {\bf 60}, 053003 (1999).


\bibitem{INO}   R. Gandhi, {\it et al.},
  Phys.\ Rev.\  D {\bf 76}, 073012 (2007).


\bibitem{Volpe} R. Lazauskas and C. Volpe, 
  Nuc. Phys. {\bf A}729, 219 (2007).


\bibitem{HaxtonTotal}   W.~C.~Haxton,
  Phys.\ Rev.\  D {\bf 36}, 2283 (1987).

\bibitem{LSNDnue}   L.~B.~Auerbach {\it et al.}  [LSND Collaboration],
  Phys.\ Rev.\  D {\bf 63}, 112001 (2001).

\bibitem{HaxtonAngle}  
 W.~C.~Haxton,
  Phys.\ Rev.\  C {\bf 37}, 2660 (1988).


\bibitem{sin2thwpaper} 
 T.~Adams {\it et al.}  [NuSOnG Collaboration],
  Int.\ J.\ Mod.\ Phys.\  A {\bf 24}, 671 (2009)


\bibitem{K2KLOI}    Y.~Ashie {\it et al.}  [Super-Kamiokande Collaboration],
  Phys.\ Rev.\  D {\bf 71}, 112005 (2005).



\end{thebibliography}
\end{document}